\documentclass[superscriptaddress,groupedaddress,nofootnoteinbib,12pt]{article}
\usepackage{color}
\usepackage{graphicx}
\usepackage{dcolumn}
\usepackage{bm}
\usepackage{amssymb}
\usepackage{amsmath}
\usepackage{sectsty}
\usepackage{colortbl}
\usepackage{latexsym}
\usepackage{float}
\usepackage{ifthen}
\usepackage{caption,subfig}
\usepackage{enumerate}
\usepackage{url}
\usepackage{caption,subfig}
\usepackage{feynmf}	
\usepackage{jcappub}

\def\ba{\begin{eqnarray}}
\def\ea{\end{eqnarray}}
\def\mpl{M_{\rm Pl}}

\def\L{\mathcal{L}}
\def\U{\mathcal{U}}
\def\K{\mathcal{K}}
\def\({\left(}
\def\){\right)}

\def\nn{\nonumber}
\def\p{\partial}
\def\mn{_{\mu \nu}}

\def\stu{St\"uckelberg }
\def\p{\partial}
\def\mupn{^\mu_{\ \nu}}
\def\<{\langle}
\def\>{\rangle}

\def\d{\mathrm{d}}

\renewcommand{\thesubsection}{\arabic{section}.\arabic{subsection}}

\begin{document}

\textit{Preprint submitted to Int. J. Mod. Phys. D $^{\copyright}$ 2014,World Scientific Publishing Company}

\title{Stable FLRW solutions in Generalized Massive Gravity}

\author{Claudia de Rham,$^{a}$ Matteo Fasiello$^{a,b}$ and Andrew J.~Tolley$^{a}$}
\affiliation{$^{a}$Department of Physics, Case Western Reserve University, \\
10900 Euclid Ave, Cleveland, OH 44106, USA}
\affiliation{$^{b}$Stanford Institute for Theoretical Physics, Department of Physics, Stanford University, Stanford, CA 94305, USA}

\emailAdd{Claudia.deRham@case.edu}
\emailAdd{Matteo.Fasiello@case.edu}
\emailAdd{Andrew.J.Tolley@case.edu}

\abstract{We present exact FLRW solutions in generalized massive gravity where the mass parameters are naturally promoted to Lorentz-invariant functions  of the \stu fields. This new dependence relaxes the constraint that would otherwise prevent massive gravity from possessing exact FLRW solutions. It does so without the need to introduce additional degrees of freedom. We find self-accelerating cosmological solutions and show that, with a mild restriction on the region of phase space, these cosmological solutions exhibit full stability, i.e. absence of ghosts and gradient instabilities for all the tensor, vector and scalar modes, for all cosmic time. We perform the full decoupling limit analysis, including vector degrees of freedom, which can be used to confirm the existence of an active Vainshtein mechanism about these solutions.
}

\maketitle

\section{Introduction}

The recent progress and development of infrared modified theories of gravity such as massive gravity has been partially motivated and fuelled by their implications for late-time cosmology \cite{deRham:2014zqa}. While a great deal of work has been done in understanding consistency issues, the exploration of their cosmological solutions has been difficult due on the one hand to the existence of many candidate cosmological solutions, and, on the other, to technical difficulties. This is particularly pertinent to the case of Massive Gravity with a Minkowski metric \cite{deRham:2010ik,deRham:2010kj}. There it can be easily shown that the constraints of massive gravity forbid the existence of FLRW solutions \cite{D'Amico:2011jj}\footnote{Although technically a background open universe solutions exists \cite{Gumrukcuoglu:2011ew}, these solutions are unstable \cite{Vakili:2012tm}.}. The absence of FLRW solutions is not, as is often claimed, a problem for massive gravity since the Vainshtein mechanism guarantees the existence of inhomogeneous/anisotropic solutions which look arbitrarily close to homogeneous over a range of scales comparable to the inverse mass of the graviton. The existence or absence of exact FLRW solutions has nothing to do with cosmological viability, observations only require the existence of solutions which are approximately homogeneous and isotropic over scales comparable to  the Hubble scale today, $H_{\rm today}^{-1}$.
Since the mass of the graviton is usually assumed to be comparable, to within a few orders of magnitude, to the Dark Energy scale, i.e.~$H_{\rm today}^{-1}$, it is possible to find solutions which are consistent with homogeneity and isotropy over most of the cosmic history \cite{Volkov:2012cf,Volkov:2012zb,Khosravi:2013axa,Gumrukcuoglu:2011zh,Gumrukcuoglu:2012aa,Gratia:2012wt,Wyman:2012iw,Kobayashi:2012fz,DeFelice:2013awa,DeFelice:2012mx,DeFelice:2013bxa,Tasinato:2013rza,Gratia:2013gka,Gratia:2013uza}. A related observation that seems to be less well appreciated is that bi-gravity models \cite{Hassan:2011zd} also admit inhomogeneous/anisotropic solutions which will look arbitrarily close to FLRW and yet have a significantly different dynamical history than the exact FLRW solutions that have been considered in the literature \cite{vonStrauss:2011mq,Comelli:2011zm,Comelli:2012db,Volkov:2011an,Akrami:2012vf,Khosravi:2012rk,Berg:2012kn,Fasiello:2013woa,DeFelice:2014nja,Comelli:2014bqa,Lagos:2014lca}. These bi-gravity solutions are the `uplift' of the associated inhomogeneous massive gravity solutions \cite{AJT:bigravity101}.
As in the case of massive gravity, it is likely that these inhomogeneous solutions are the correct solutions to look for cosmologically viable bi-gravity models. Indeed we expect by causality arguments, which are at the origin of the horizon problem, that the Universe is in fact inhomogeneous at scales larger than the current Hubble radius.
\\

The additional degrees of freedom in massive gravity and bi-gravity, relative to General Relativity (GR), mean that many such solutions exist and so there are many candidate cosmologies. Unfortunately it is difficult to explore the space of possibilities analytically. The majority of these solutions will be ghost-free and stable and exhibit the correct number of propagating degrees of freedom \cite{deRham:2010tw,Motloch:2014nwa}. A number of exact solutions are known \cite{Volkov:2013roa}, but most of these are in branches leading to ghostly pathologies and therefore do not lie in the regime of validity of the EFT (for a recent discussion that resolves a number of these issues see \cite{Motloch:2014nwa}). The existence of most of the exact solutions found so far can be traced to choosing non-standard branches in the equations of motion, which are typically easier to solve, and which is also the underlying reason for the instabilities.
\\

From a calculational point of view alone, it is simpler if exact FLRW solutions can be found, since perturbations can then be analyzed in the usual way based on the representations of the group of isometries. With this in mind, many authors have looked at extensions to massive gravity which introduce additional degrees of freedom, beyond the five of the standard massive graviton. These examples include bi-gravity \cite{Hassan:2011zd,vonStrauss:2011mq,Comelli:2011zm,Comelli:2012db,Volkov:2011an,Akrami:2012vf,Khosravi:2012rk,Berg:2012kn,Fasiello:2013woa,DeFelice:2014nja,Lagos:2014lca}, the Quasi-Dilaton model \cite{D'Amico:2012zv,D'Amico:2013kya} and its generalizations \cite{Mukohyama:2013raa,DeFelice:2013dua,Comelli:2014bqa,DeFelice:2013tsa}, mass varying gravity \cite{Huang:2012pe,Leon:2013qh,Wu:2013ii,Motohashi:2014una}, multi-vierbein gravity \cite{Hinterbichler:2012cn}, extended massive gravity \cite{Hinterbichler:2013dv,Andrews:2013uca}. A recent approach makes use of the non-minimal coupling of matter which although leads to a ghost, the mass of the ghost is above the strong coupling scale \cite{deRham:2014naa,Gumrukcuoglu:2014xba,deRham:2014fha}. This requires the existence of a field which couples non-minimally to the metric (this field could be a `dark sector' degree of freedom or the inflaton, etc\ldots, or would require that matter and radiation couple differently). This possibility is logically distinct from the approach we follow here. \\

While these are certainly worthwhile avenues to explore, a genuine concern arises that the predictions of the model will then dependent strongly on the new degrees of freedom one introduces, and have little to do with the properties of the massive graviton itself.
With this in mind, it is interesting to explore a modification of massive gravity that admits FLRW solutions but does not introduce any new dynamical degrees of freedom. One such approach is to consider massive gravity on an FLRW reference metric, for instance de Sitter \cite{deRham:2012kf,Fasiello:2012rw}. However this model was shown to exhibit problems simultaneously satisfying stability and observational viability. This result was unfortunate since massive gravity on de Sitter has the same number of symmetries and aesthetic appeal as massive gravity on Minkowski. It would be interesting to look for simple alternatives that have the character of maintaining a large degree of symmetry, without introducing new degrees of freedom. \\

In this article we will explore the cosmology of precisely such a model that was recently proposed in \cite{deRham:2014lqa} within the context of Galileon dualities \cite{deRham:2013hsa}. This model preserves the global Lorentz symmetry of the original massive gravity Lagrangian and introduces no new degrees of freedom. Massive gravity on Minkowski, when written in `St\"uckelberg--ised' form, admits a local ${\rm Diff}(M)$ gauge symmetry, and a global Poincar\'e symmetry. The latter global symmetry is the isometry group of the reference metric. This isometry group is preserved by the vacuum of the theory which was inherent in its construction as a theory of an interacting massive spin-2 representation of the Poincar\'e group. Since the latter symmetry is global, a natural way to modify the theory that does not run afoul of new degrees of freedom is to break the Poincar\'e symmetry down to a subgroup. There are a number of ways to do this leading to substantively different theories. One possibility is to retain translation invariance, but break Lorentz invariance. This gives rise to Lorentz violating massive gravity theories which have been considered at length in the past \cite{Rubakov:2004eb,Dubovsky:2004sg} (for a review see \cite{Rubakov:2008nh}) and in the context of more recent developments as ghost-free models still propagating 5 degrees of freedom in \cite{Comelli:2013txa,Comelli:2013tja,Lin:2013aha,Langlois:2014jba,Comelli:2014xga}). The other logical extreme is to maintain Lorentz invariance but break translation invariance. This is achieved simply by allowing the parameters in the massive gravity Lagrangian to become dependent on the Lorentz invariant combination of \stu fields $\phi^a \phi_a$. It is this later extension that we will refer to as `Generalized Massive Gravity'\footnote{For the context of this paper we refer to `Generalized Massive Gravity', a theory of massive gravity where the mass parameters are promoted to the Lorentz invariant combination of the \stu fields. This is distinct from promoting the mass parameters to a function of another external scalar field with its own kinetic term as considered in \cite{Huang:2013mha}. The theory considered in this paper is also distinct from the theory proposed in Ref.~\cite{Bergshoeff:2009hq} which is also sometimes referred to as Generalized Massive Gravity. } in what follows. In \cite{deRham:2014lqa} it was argued that this theory would necessarily propagate only the 5 original degrees of freedom of the massive graviton. Subsequently closely related generalizations that preserve SO(3) but allow the mass parameters to depend arbitrarily on $\phi^0$ were considered in \cite{Langlois:2014jba,Comelli:2014xga}. We will see that in fact by taking an appropriately defined `zero curvature scaling limit' our cosmological solutions will be related to the ones considered in \cite{Langlois:2014jba}.
\\

The Lorentz invariant `Generalized Massive Gravity' theory considered in \cite{deRham:2014lqa} was arrived at by a rather different means. Although it has been known for some time that the Galileon Lagrangians \cite{Nicolis:2008in} have a natural origin in the decoupling limit of massive gravity \cite{deRham:2009rm,deRham:2010gu,deRham:2012az}, the so-called generalized Galileons \cite{Deffayet:2009mn} did not. The `Generalized Massive Gravity' theories are precisely those theories which provide a gravitational embedding of the generalized Galileon\footnote{Although it is possible to covariantize the generalized Galileon Lagrangian, an operation which results in the Horndeski models, this covariantization treats $\phi$ as a spin-0 field which is disconnected from the graviton multiplet.} in which the Galileon field $\pi$ continues to play its role as the helicity-0 mode of the graviton. These models are then in a sense natural covariantizations, or more precisely IR completions of the Minkowski space generalized Galileons. \\

In what follows we will show that these Generalized Massive Gravity theories admit fully stable cosmological solutions (no ghosts, or gradient instabilities in the tensor, vector or scalar sector). Furthermore we show that these theories admit self-accelerating solutions. We shall perform the analysis of perturbations in the decoupling limit (DL) of the theory, however the FLRW background solutions are exact and therefore there is no difficulty in extending the stable DL solutions to stable exact solutions of the theory. The decoupling limit analysis also allows us to see that there is an active Vainshtein mechanism in place, whose role will be important in the growth of non-linear perturbations and the observational viability of this theory. We also take care to include the contribution from the non-zero background for vectors in the DL which had been overlooked in most previous related analyses.\\

{\it Outline:} The rest of the paper is organized as follows: In section~\ref{sec:FLRW sol} we give a brief overview of the Generalized theory of massive gravity. We show how to obtain cosmological solutions out of this manifestly Lorentz invariant theory and present a class of exact FLRW solutions. We also show the existence of self-accelerating solutions which could {\it a priori} have a viable cosmic history.  In section~\ref{sec:DL} we then derive the decoupling limit of the theory and recover the exact FLRW solutions found in the full theory. This illustrates the power of the decoupling limit. We then use this decoupling limit to analyze the stability of these FLRW solutions in section~\ref{dec:stability}. In doing so we take great care of the background contributions from all the modes including the vector ones. A peculiarity of this generalized theory is a linear mixing between the helicity-1 and -0 modes which has to be diagonalized. We find that all the helicity-2,-1 and -0 modes can be stable (free of ghosts and gradient instabilities) for all cosmic time, provided the parameters and functions of the generalized theory satisfy an acceptable set of conditions. We summarize our results in section~\ref{sec:Discussion} and present open avenues. Appendix~\ref{Appendix} provides a vierbein derivation of the decoupling limit.

\section{Exact FLRW solutions in Generalized Massive Gravity}
\label{sec:FLRW sol}

We begin with the Generalized Massive Gravity Lagrangian considered in \cite{deRham:2014lqa}. To reiterate this Lagrangian preserves the global Lorentz invariance of the original massive gravity Lagrangian \cite{deRham:2010kj} but breaks the translation invariance $\tilde \phi^a \rightarrow \tilde \phi^a +c^a$. This breaking is seen explicitly through the mass terms becoming dependent on the Lorentz invariant combination $\tilde \phi^a \tilde \phi_a=\eta_{ab}\tilde \phi^a \tilde \phi^b$.
\ba
\label{eq:GMG}
S=\frac{\mpl^2}{2} \int \d^4 x \sqrt{-g}\left[
 R[g]+\frac{m^2}{2} \sum_{n=0}^4 \tilde \alpha_n (\tilde \phi^a \tilde \phi_a)\,  \mathcal{U}_n[\K]
\right]+\mathcal{L}_{\rm matter}[g, \psi^{(i)}]\,,
\ea
where we use the same convention as in \cite{deRham:2014zqa} where the potential terms are given symbolically by $\U_n[\K]= \mathcal{E}\mathcal{E}\K^n = (4-n)! [\K]^n+ \cdots$.
The $\tilde\phi^a$ are the \stu fields and we work with the flat reference metric so that
\ba
&& \tilde \phi^a  \tilde \phi_a =  \eta_{ab}  \tilde \phi^a \tilde \phi^b \, ,\\
&& \K\mupn = \delta\mupn- \(\sqrt{g^{-1}f}\)\mupn \, ,\\
&& f\mn = \p_\mu \tilde \phi^a \p_\nu \tilde \phi^b \eta_{ab}\,,
\ea
and the global Lorentz symmetry is manifest through contractions with the Minkowski metric $\eta_{ab}$. We will also use the equivalent representation \cite{Hassan:2011vm}
\ba
\label{eq:GMGa}
S=\frac{\mpl^2}{2} \int \d^4 x \sqrt{-g}\left[ R[g]-\frac{m^2}{2} \sum_{n=0}^4  \frac{\tilde \beta_n(\tilde \phi^a \tilde \phi_a)}{n!} \,  \mathcal{U}_n[\sqrt{g^{-1}f}] \right]+\mathcal{L}_{\rm matter}[g, \psi^{(i)}] \, ,
\ea
for which
\ba
\label{eq:betatilde}
\tilde \beta_k (\tilde \phi^a\tilde \phi_a)= (-1)^{k+1}\sum_{n=0}^4 \frac{n!}{(n-k)!} \tilde \alpha_n(\tilde \phi^a\tilde \phi_a)\,.
\ea

\subsection{From manifest Lorentz invariance to Cosmology}

The cosmological solutions we look for will preserve the global Lorentz symmetry of the original theory. This is achieved by looking at open universe solutions, for which the Lorentz symmetry $SO(1,3) $ acts as the isometry group of the open spatial slices $H^3$. We can make this manifest by choosing a version of `unitary gauge' in which the \stu fields describe the open slicing $k=-|k|$ of Minkowski space-time (see for instance \cite{Gumrukcuoglu:2011ew})
\ba
\tilde \phi^0 = f(t) \sqrt{1+ |k| \vec x\, ^2} \, , \quad  \tilde \phi^i = \sqrt{|k|} f(t) x^i \, .
\ea
A short calculation shows that in  `almost' unitary gauge
\ba
 \tilde \phi^a \tilde \phi_a = -f(t)^2 \, , \quad  f_{\mu \nu} \d x^{\mu} \d x^{\nu} = - \dot f(t)^2 \d t^2 +  |k| f^2 \,  \d \Omega^2_{H^3}\, ,
\ea
where we have defined the metric on $H^3$ as
\ba
\d \Omega^2_{H^3} = (\d \vec x)^2 - \frac{|k|  }{1+|k| \vec x\, ^2} (\vec x .\d \vec x)^2 \, ,
\ea
which makes manifest the FLRW open slices of Minkowski.  Although there is no difficulty is maintaining the spatial curvature $k$, the calculations of perturbations in the decoupling limit are more easily performed by first taking a scaling limit $|k| \rightarrow 0$, (holding $\sqrt{|k|}f$ fixed). We will refer to this as the {\it zero spatial curvature scaling limit}. The action remains finite in this scaling limit, and no degrees of freedom are lost. Furthermore we shall see that the perturbations can remain stable in this limit and so the curvature $k$ is largely irrelevant to the dynamics of the system. In order to take the limit we first perform a redefinition of the function $f(t)$ as
\ba
f(t) = \frac{1}{\sqrt{|k|}} + \chi(t) \, ,
\ea
so that the limits may be take as follows
\ba
&& \lim_{|k| \rightarrow 0} \frac{\sqrt{|k|}}{2}\left( \tilde \phi^a \tilde \phi_a + \frac{1}{|k|} \right)  \rightarrow -\chi(t) \, , \\
&& \lim_{|k| \rightarrow 0} f_{\mu \nu} \d x^{\mu} \d x^{\nu}  \rightarrow - \dot \chi(t)^2 \d t^2  + (\d \vec x)^2 \, .
\ea
Finally we make the following redefinition of the mass function parameters
\ba
\tilde \alpha_n ( \tilde \phi^a  \tilde \phi_a) = \alpha_n\left(- \frac{ \sqrt{|k|}}{2}\left(\tilde  \phi^a \tilde \phi_a + \frac{1}{|k|} \right)  \right) \, .
\ea
This redefinition, which may always be performed, is such that in unitary gauge and in the zero spatial curvature scaling limit we have
\ba
\lim_{|k| \rightarrow 0}  \tilde \alpha_n ( \tilde \phi^a \tilde \phi_a)  =  \alpha_n ( \chi(t))  \, .
\ea
Given these simplifications, we shall find it convenient to work with the $\alpha_n(\chi)$ and define a new set of \stu fields $\phi^a$ for which in this gauge $\phi^0 = \chi(t)$ and $\phi^i = x^i$.

\subsection{Zero spatial curvature scaling limit and Decoupling limit}

The `zero spatial curvature scaling limit' may be performed at the level of the action and results in the following effective theory
\ba
\label{eq:GMG2}
S=\frac{\mpl^2}{2} \int \d^4 x \sqrt{-g}\left[
 R[g]+\frac{m^2}{2} \sum_{n=0}^4  \alpha_n (\phi^0)\,  \mathcal{U}_n[\K]
\right]+\mathcal{L}_{\rm matter}[g, \psi^{(i)}]\,.
\ea
The $\phi^a$ are the redefined \stu fields associated with the `almost' unitary gauge $ \phi^0 = \chi(t)$ and $\phi^i = x^i$ (in which we maintain time reparameterization invariance through $\chi(t) $). $\K$ has the same definition as before
\ba
 \K\mupn = \delta\mupn- \(\sqrt{g^{-1}f}\)\mupn \, ,
\ea
whereas now the reference metric is Minkowski in flat slicing
\ba
 f\mn = \p_\mu  \phi^a \p_\nu  \phi^b \eta_{ab} = -\dot \chi(t)^2 \d t^2 + \d \vec x^2  \quad \text{(in `almost' unitary gauge)}\,.
\ea
This is simply standard massive gravity for Minkowski space-time with time-dependent mass parameters. This effective action and its generalizations have been considered in two works \cite{Langlois:2014jba,Comelli:2014xga} and shown to be free of the BD ghost independently of the Generalized Massive Gravity analysis performed in \cite{deRham:2014lqa}. These results are of course consistent as guaranteed by the scaling limit argument. The authors of \cite{Langlois:2014jba,Comelli:2014xga}  allow a somewhat more general form since their models explicitly break SO(3,1) invariance. By contrast our starting Lagrangian is manifestly SO(3,1) invariant and this provides a stronger restriction on the form of the Lagrangian.

To reiterate, in this form, it may appear as if we had broken Lorentz invariance by picking a preferred time direction, however as emphasized earlier, it is crucial for our analysis that this is obtained as a scaling limit of a {\bf Lorentz invariant theory}. This is consistent since the Lorentz symmetry acts on the spatial hyperboloid and leaves $t$ invariant. In the $k \rightarrow 0$ limit the Lorentz symmetry degenerates into the $ISO(3)$ isometry group of $R^3$ via a In\"on\"u-Wigner contraction.\\

Alternatively, one could also view the effective theory \eqref{eq:GMG2} as a small distance limit of large distance inhomogeneities and this effective action would therefore nicely fit in the greater picture of massive gravity where we expect all the solutions to be inhomogeneous at large distances (compared to the size of the current observable Universe).  This is of course also what is expected from a cosmological viewpoint. \\

Our analysis of perturbations will be done in the standard massive gravity decoupling limit. This analysis will thus be complementary to the discussion in \cite{Langlois:2014jba} with the distinction being that the decoupling limit maintains the most important non-linearities which are important for demonstrating the existence of a Vainshtein mechanism. On the other hand the decoupling limit analysis is valid only at sub-horizon scales and so is immune to super-horizon instabilities. However super-horizon instabilities at the Hubble scalar are harmless and so these do not pose a serious concern\footnote{In a local field theory, the stability of a super-horizon mode for which $|\vec k| \ll a H$ is essentially equivalent to the stability of the zero mode since locality requires that no terms blow up as inverse powers of $| \vec k|$. This is the essence of the {\it Separate Universe} idea. This may be analyzed by working with the mini-superspace Lagrangian and analyzing the stability of the general aniostropic, spatially curved cosmology. }.\\

The decoupling limit is introduced as usual by writing the $ \phi^a $ \stu fields (not the $\tilde \phi^a$) in a vector/scalar decomposition about unitary gauge $\chi(t)=t$:
\ba
 \phi^a = x^a - \frac{V^a}{m \mpl} - \frac{\eta^{ab} \partial_b \pi}{\Lambda^3} =  x^a - m \frac{V^a}{\Lambda^3} - \frac{\eta^{ab} \partial_b \pi}{\Lambda^3}   \, .
\ea
In particular we see that at leading order in the decoupling limit $\phi^0 = t+ \frac{\dot \pi}{\Lambda^3}$ and so the mass parameters may be taken to be of the form
\ba
\tilde \alpha_n (\tilde \phi^a \tilde \phi_a)\to  \alpha_n(t+\frac{\dot \pi}{\Lambda^3})\,.
\ea
At a pragmatic level this means that the new feature of the DL derivation will be new interactions coming from additional $\dot \pi$ terms which do not normally arise. These new interactions will break Lorentz invariance (from the perspective of transformations on $(x,t)$ - not the original Lorentz invariance of the theory), but not rotational invariance. We will also find additional mixing terms coming from the scalar and vector which do not normally arise.

\subsection{Exact FLRW solutions and alleviating the FLRW constraint}

\subsubsection{Open-slicing FLRW solutions}

We now come to the central point which is the demonstration of the existence of exact FLRW solutions.
We start with the dynamical and reference metric in the open slicing FLRW form
\ba
\d s_g^2 &=&  - N(t)^2 \d t^2 +  a(t)^2 \,  \d \Omega^2_{H^3} \, , \\
\d s_f^2 &=&  - \dot f(t)^2 \d t^2 +  |k| f(t)^2 \, \d \Omega^2_{H^3} \, .
\ea
Including the energy density $\rho(a(t))$ of a minimally coupled matter field, the mini-superspace action takes the form
\ba
\label{eq:mini1a}
&& S_{\text{mini-superspace}}= \int \d t \,\Bigg\{ - 3 \mpl^2 \frac{a \dot a^2}{N} + 3 \mpl^2 a^3 N \frac{k}{a^2} -\rho(a) N a^3 \\
&& +\frac 32 m^2 \mpl^2 N a^3
\sum_{n=0}^4 \left[ (4-n)\(1-\sqrt{|k|} \frac{f}{a}\)^n  + n \(1-\sqrt{|k|} \frac{f}{a}\)^{n-1}\(1-\frac{\dot f}{N} \)\right] \tilde \alpha_n(-f^2) \Bigg\} \,.\nn
\ea
In this form the action preserves time reparameterization invariance from which it follows that the $\ddot a$ `Raychaudhuri' equation is not independent of the $\dot a$ Friedmann constraint equation, and so may be ignored.
As usual, varying with respect to the lapse leads to the Friedmann equation for the Hubble parameter $H=\dot a /(a N)$
\ba
\label{eq:Friedmann1a}
3\mpl^2 \( H^2+ \frac{k}{a^2} \) &=& \rho-\frac 32 m^2 \mpl^2 \sum_{n=0}^4\left[(4-n) \(1-\sqrt{|k|} \frac{f}{a} \)^n+n \(1-\sqrt{|k|} \frac{f}{a} \)^{n-1}\right] \tilde \alpha_n(-f^2) \nn \\
&=& \rho+\frac{m^2\mpl^2}{4}\left[24 \tilde \beta_0+18 \frac{\tilde \beta_1 \sqrt{|k|} f}{a}+6 \frac{\tilde \beta_2  |k| f^2}{a^2}+\frac{\tilde \beta_3 |k|^{3/2} f^3 }{a^3}\right]
\,,
\ea
remembering that the $\tilde \beta_n$ functions are defined as in \eqref{eq:betatilde}.

To close the Friedmann equation we need to determine $f(t)$,  which is fixed by the following non-dynamical equation
\ba
\label{eq:constraint1a}
&& m^2 H \left[\frac 32   \sqrt{|k|} \tilde \beta_1 a^2+ \tilde \beta_2a |k| f +\frac 14 \tilde \beta_3 |k|^{3/2} f^2 \right]
=  \nn\\
&& \frac{m^2 }{2 a}\left[4  \frac{\partial \tilde \beta_0}{\partial f} a^3+3 f \frac{\partial \tilde \beta_1}{\partial f}  \sqrt{|k|} a^2+f ^2\frac{\partial  \tilde \beta_2}{\partial f} |k|  a+\frac 16 f^3 \frac{\partial \tilde \beta_3}{\partial f} |k|^{3/2}  \right]\,.
\ea
Despite its cumbersome appearance, this is a formally straightforward equation to solve which determines $f$ in terms of $H$ and $a$. The solution may then be substituted back into (\ref{eq:Friedmann1a}) to give a closed solvable Friedmann equation\footnote{The system is closed since we may use time-reparameterization invariance to gauge fix the lapse $N$, e.g. as $N=1$.}.

\subsubsection{Zero-curvature FLRW solutions}

In what remains we shall for simplicity work in the zero spatial curvature scaling limit theory defined by the action (\ref{eq:GMG2}), bearing in mind that these solutions are in one to one correspondence with exact open universe FLRW solutions in the full theory (\ref{eq:GMG}) as we demonstrate below. In taking this limit our subsequent decoupling limit analysis will also apply for specific cases considered in \cite{Langlois:2014jba,Comelli:2014xga}.
To reiterate the steps in this limit we start with the dynamical metric in the FLRW form and maintain the time reparameterization invariance implied by the choice of \stu fields $ \phi^0 (t) =\chi(t)$ and $\phi^i = x^i$,
\ba
\d s_g^2&=& -N^2(t) \d t^2+ a^2(t)\d \vec x\, ^2 \, ,\\
\d s_f^2 &=& -\dot \chi^2(t) \d t^2+ \d \vec x\, ^2\,,
\ea
and similarly defining the linear combinations of the mass functions
\ba
\label{eq:beta}
\beta_k(\chi)= (-1)^{k+1}\sum_{n=0}^4 \frac{n!}{(n-k)!} \alpha_n(\chi)\,.
\ea
We obtain the mini-superspace action
\ba
S_{\text{mini-superspace}}&=& \int \d t \,\Bigg\{ - 3 \mpl^2 \frac{a \dot a^2}{N}-\rho(a) N a^3 \\
&+&\frac 32 m^2 \mpl^2 N a^3
\sum_{n=0}^4 \left[ (4-n)(1-a^{-1})^n  + n(1-a^{-1})^{n-1}\(1-\frac{\dot \chi}{N} \)\right] \alpha_n(\chi) \Bigg\} \,.\nn
\ea
The resulting Friedmann equation takes the form
\ba
\label{eq:Friedmann1}
3\mpl^2 H^2 &=& \rho-\frac 32 m^2 \mpl^2 \sum_{n=0}^4\left[(4-n)(1-a^{-1})^n+n (1-a^{-1})^{n-1}\right] \alpha_n(\chi)\\
&=& \rho+\frac{m^2\mpl^2}{4}\left[24 \beta_0(\chi)+18 \frac{\beta_1(\chi)}{a}+6 \frac{\beta_2(\chi)}{a^2}+\frac{\beta_3(\chi)}{a^3}\right]
\,,
\ea
and the constraint equation that determines $\chi$ is
\ba
\label{eq:ConstraintH}
m^2 H \left[\frac 32 \beta_1 a^2+\beta_2a +\frac 14 \beta_3 \right]
= \frac{m^2 }{2 a}\left[4\beta_0'a^3+3 \beta_1'a^2+\beta_2'a+\frac 16 \beta_3'\right]\,,
\ea
where it is implicit that all the functions $\beta_n$ depend on $\chi(t)$ and $\beta_n' = \partial \beta_n/\partial \chi$.
These equations are equivalently obtained as the $k \rightarrow 0$ scaling limit of the previous set (\ref{eq:mini1a}), (\ref{eq:Friedmann1a}), (\ref{eq:constraint1a}), remembering that
\ba
\sqrt{|k|} f \rightarrow 1 \, , \quad \, f \frac{\partial \tilde \beta_n}{\partial f} \rightarrow \frac{\partial \beta_n }{\partial \chi} \, .
\ea
Since $\chi$ enters as an auxiliary variable, then as before this equation determines the evolution of the \stu field in terms of $H$ and $a$.

\subsubsection{Constraint in standard massive gravity}

In standard massive gravity \cite{deRham:2010kj} the $\alpha_n$ are constant and so are the $\beta_n$. So in the standard massive gravity case,  \eqref{eq:ConstraintH} reduces to
$m^2 \left[\frac 32 \beta_1 a^2+\beta_2a +\frac 14 \beta_3 \right]\dot a = 0$,
which is nothing other than the  well-known constraint which forbids the existence of exact (spatially flat or closed) FLRW solutions in massive gravity\footnote{Note however that this does not forbid this existence of solutions which are arbitrarily close to FLRW within our horizon and only start exhibiting inhomogeneities or anisotropies on scales larger than the observable Universe \cite{D'Amico:2011jj}.} \cite{D'Amico:2011jj}. For generalized massive gravity \eqref{eq:GMG}, as soon as the coefficients $\alpha_n$ are promoted to functions of the \stu fields, we see that this constraint no longer forbids the existence of exact FLRW in massive gravity. Rather one can read this equation~\eqref{eq:ConstraintH} as an implicit relation for the \stu field $\chi(t)$ in terms of $H$ which can then be substituted into the Friedmann equation \eqref{eq:Friedmann1}. This is consistent with how the equation is obtained in the full theory as the equation of motion for the $\phi^0$ \stu field. As we shall in the decoupling limit Eq.~\eqref{eq:ConstraintH} corresponds to the equation of motion for $\pi$.

\subsection{Self-accelerating solutions}

\label{sec:SA}

The above Friedmann equation \eqref{eq:Friedmann1} contains a rich family of solutions given the specification of the free functions $\beta_n(\chi)$. However the stability analysis we perform later will be restricted to the massive gravity decoupling limit. For this reason we shall concentrate on the form of the solutions which remain well defined in the decoupling limit. We shall see later that the functions that are well defined in this limit take the form
\ba
\beta_n(\chi) = \bar \beta_{n} +  \bar \beta_{n,1}\ m \chi +\frac{m^2}{\Lambda^2} \beta_{n,2}(\chi) \, ,
\ea
where the bar quantities $\bar \beta_n$ and $\bar{\beta}_{n,1}$ are constant. The
 $\beta_{n,2}(\chi)$ are arbitrary functions of $\chi$ which are assumed to remain finite in the DL. 
 We recall that $\chi$ has dimension $[{\rm mass}]^{-1}$ and so the parameters $\bar \beta_n$, $\bar \beta_{n,1}$ and the functions $\beta_{n,2}$ are all dimensionless. 
 As we shall see later, this ansatz arises from the fact that in the DL the terms $ \bar \beta_{n} $ and $ \bar \beta_{n,1}$ lead to total derivatives at leading order and so may be taken as finite, whereas the $\beta_{n,2}$ function gives rise to an interaction which would diverge were it not for the $m^2$ in front. Given this in the DL $m\rightarrow 0$, the $\beta_{n,2}$ do not enter the Friedmann equation, although they do enter the stability analysis. As we shall see in section \ref{sec:DL}, in the DL the Hubble parameter scales as $m$. \\

For convenience we also define the following functions of the scale factor,
\ba
\label{eq:Sigma}
\Sigma_{k,i} (a) =\sum_{n=0}^4 \frac{(4-n)}{(n-k)!}\beta_{n}^{(i)}(\chi)\ a^{3-n}\qquad
{\rm and}\qquad
S_{k,i} (a) =\sum_{n=0}^4 \frac{(4-k)}{(n-k)!}\beta_{n}^{(i)}(\chi)\ a^{3-n}\,,
\ea
where we use the notation $\beta_{n}^{(0)}(\chi)=\bar \beta_n$, $\beta_{n}^{(1)}(\chi)=\bar \beta_{n,1}$
and $\beta_{n}^{(2)}(\chi)=\beta_{n,2}''(\chi)$. All the $\Sigma_{k,i}(a)$ and $S_{k,i}(a)$ are manifestly positive for any $k=0,\cdots,4$ if all the $\beta_{n,i}$ are positive for $n=0,\cdots, 4$.

With this in mind let us consider the cosmology of the special case $\beta_n(\chi) = \bar \beta_{n} +  \bar \beta_{n,1} \ m \chi $. We should first solve the constraint equation (\ref{eq:ConstraintH}) to determine $\chi$. The equation is
\ba
 m^2 H \left[\Sigma_{1,0} + m \chi \Sigma_{1,1}  \right]  =\frac{m^3 }{a}\Sigma_{0,1}\,,
\ea
which gives on rearrangement
\ba
\chi=\frac{m \Sigma_{0,1}- a H \Sigma_{1,0}}{m a H \Sigma_{1,1}}\,.
\ea
The analogue to this equation in the DL is the equation for $\pi$. Since $\chi = t+ \dot \pi/\Lambda^3$ we obtain not this equation but its time-derivative. Although the equation for $\chi $ appears to blow up as $m\rightarrow 0$, the relevant equation is that for $\dot \chi = 1+ \ddot \pi/\Lambda^3$ and since the time derivative gives an extra factor of $H$ which scales as $m$ in the DL, $H\sim m$,  it follows that this equation remains indeed finite in the DL.  \\

Substituting the solution for $\chi$ back into the Friedmann equation we obtain
\ba
\label{eq:Friedmann2}
 3\mpl^2 H^2= \rho+\frac{3}{2}\frac{m^2\mpl^2}{a^3}\Bigg[
 \Sigma_{0,0}+\frac{\Sigma_{0,1}}{\Sigma_{1,1}}\(\frac{m}{aH}\Sigma_{0,1}-\frac 12 \Sigma_{1,0}\)  \Bigg]\,. \qquad
\ea
The usual cosmological constant term is contained in $\bar \beta_{0}$ that enters in $\Sigma_{0,0}$. Let us consider this to be zero to see if there are self-accelerating solutions. To simplify the problem let us further assume $\bar \beta_{n} \approx 0$, except for $\bar \beta_2$ but the remaining coefficients are non-zero (it is necessary to keep one of the $\bar \beta_n$ nonzero to have well-defined perturbations). With this assumption at late times, i.e. large $a(t)$, the dominant contribution to the Friedmann equation is
\ba
3\mpl^2 H^2  \approx \rho  +8 m^2\mpl^2 \frac{ \bar \beta_{0,1}^2}{\bar \beta_{1,1}} \frac{m}{  H }+ \dots \, .
\ea
The qualitative form of this equation is that at early times $\rho$ dominates and we recover the usual GR expansion (with additional `dark' curvature components), whereas at late times the second term kicks in and leads to a period of self-acceleration for which $H \rightarrow $ constant, which corresponds to an effective dark energy equation of state $w \rightarrow -1$. The general case will be more subtle and will lead to an effectively dynamical dark energy.

Notice that these types of theories typically enjoy a technically natural mass parameter $m$ and coefficients $\bar \beta_{n,1}$ \cite{deRham:2012ew,deRham:2013qqa} (see also \cite{deRham:2014wfa} for related discussions). This type of theories therefore represent potentially good candidates to tackle the new Cosmological Constant problem. \\

To make this more explicit let us consider the case where the only non-zero coefficients are $\bar \beta_{0,1}$ and $\bar \beta_{1,1}$ and $\bar \beta_2$. Then the exact Friedmann equation (in the $k=0$ limit) is
\ba
\label{eq:Friedmann2}
 3\mpl^2 H^2 = \rho+ \frac{m^3\mpl^2}{2 H  }\left[   \frac{\left(4\bar \beta_{0,1}+3 \bar \beta_{1,1}a^{-1} \right)^2    }{ \bar \beta_{1,1}   }    \right] - 2 m^2 \mpl^2 \frac{\bar \beta_{0,1} \bar \beta_2}{a \bar \beta_{1,1}}
\,. 
\ea
Although we have tuned to a special set of parameters, this form nicely illustrates the fact that the Generalized massive gravity model can happily accommodate plausible self-accelerating solutions.
Furthermore by rearranging this equation in the form
\ba
\label{eq:Friedmann2a}
3\frac{ H^2}{m^2} = \frac{\rho}{m^2 \mpl^2} +\frac{m}{2 H  }\left[   \frac{\left(4\bar \beta_{0,1}+3 \bar \beta_{1,1}a^{-1} \right)^2    }{ \bar \beta_{1,1}   }    \right]- 2 \frac{\bar \beta_{0,1} \bar \beta_2}{a \bar \beta_{1,1}}\,, 
\ea
and remembering that in the DL both $H/m$, and $\rho/(m^2 \mpl^2)$ are kept fixed in the scaling limit (they still depend on time, the assumption is that the magnitude of $H$ is reduced at the same rate as $m$ at all times), we see that this Friedmann equation remains finite in the decoupling limit. This means that the stability analysis we perform will be valid for these cosmological solutions.

\section{Derivation of Decoupling Limit}

\label{sec:DL}

The standard DL of massive gravity is defined by taking the limit $m \rightarrow 0$ and $\mpl \rightarrow \infty$ keeping the strong coupling scale $(m^2\mpl)^{1/3}\equiv \Lambda $ constant. This limit is designed to focus on the scale of the leading gravitation interactions of the theory which arise principally from the helicity-zero mode. There are also additionally interactions between the vectors and the helicity-zero mode which were first given in full form in \cite{Ondo:2013wka} (see also \cite{Gabadadze:2013ria}). In the usual case the mass parameters (which are defined with an overall $m^2$ in front) are kept finite in the decoupling limit. Since curvature scales as $R \sim \mpl^{-1} \sim m^2$ in this limit, then the Hubble parameter $H$ scales as $m$, i.e. the DL can be taken on the Friedmann equation maintaining $H/m$ fixed. A minor rearrangement of the Friedmann equation shows that all the terms survive in this limit
\ba
\label{eq:Friedmann}
3 \frac{H^2}{m^2}= \frac{\rho}{m^2 \mpl^2}
+\frac 3{2a^3} \Sigma_{0,0}\,.
\ea
since we scale $\rho/m^2 \mpl^2$ as fixed. This is consistent with all known previous cases where it was possible to see the form of the Friedmann equation already in the decoupling limit (see for instance \cite{Fasiello:2013woa} where the DL is derived for massive gravity with general reference metrics and for bi-gravity).

\subsection{Scaling of the parameters}

In the present case we must also specify the scaling of the $\chi$ dependence of the functions $\alpha_n(\chi)$, or $\beta(\chi)$ which determines the rate of variation of the mass parameters. Inspection of the action shows that the mass terms have a pre-factor $\mpl^2 m^2 = \Lambda^3 \mpl = \Lambda^6/m^2$.  This looks like it is blowing up as $m\rightarrow 0$, however we are saved by the fact that the characteristic polynomial structure ${\cal U}_n[K]$ are at leading order total derivatives in the DL. When the mass parameters are functions of space and time this is no longer true, to see what happens suppose we perform an expansion in powers of the argument
\ba
\tilde \alpha_n (\phi^a \phi_a)\to \alpha_n(t+\frac{\dot \pi}{\Lambda^3})= \sum_{k\ge 0} \frac{1 }{k!}\(t+\frac{\dot \pi}{\Lambda^3}\)^k \alpha^{(k)}_n(0) \,.
\ea
Consistent with the usual case we find that the leading $\alpha_n^{0}(0)$ terms are total derivatives and so $\alpha_n^{0}(0)$ may be kept finite in the DL. Remarkably we also find that the terms coming from $\alpha_n^{1}(0)$ are also total derivatives at leading order for the pure $\pi$ interactions. However the $\alpha_n^{1}(0)$ do lead to a mixing between the vectors and scalars which only remains finite if we assume $\alpha_n^{1}(0)$ scales as $m$.
Finally for $n\ge 2$ the leading terms are not total derivatives and so for the action to remain finite in the decoupling limit we must assume that all these terms scale as $m^2$ to cancel the $1/m^2$ in the $\mpl^2 m^2 = \Lambda^3 \mpl = \Lambda^6/m^2$ prefactor. \\

In summary the full DL scaling which we shall use in the following is
\ba
m  \to 0\,, \qquad \mpl \to \infty\,, \quad {\rm keeping}\quad (m^2\mpl)^{1/3}\equiv \Lambda \to {\rm fixed}\,, \quad T^{\mu\nu}/\mpl \to {\rm fixed},
\ea
combined with the assumption that the mass parameter functions have the generic form
\ba
\alpha_n(\chi) = \bar \alpha_n + \bar  \alpha_{n,1} \ m \chi +\frac{m^2}{\Lambda^2} \alpha_{n,2}(\chi) \, ,
\ea
where
$\bar \alpha_n$ and $\bar \alpha_{n,1}$ are constants, whereas $\alpha_{n,2}(\chi)$ remain arbitrary functions. We will make use of the relations \eqref{eq:Sigma}  for the functions $\Sigma_{n,i}$ given in terms of the $\beta_n$'s
\ba
\beta_n(\chi) = \bar \beta_n +  \bar \beta_{n,1}\ m \chi + \frac{m^2}{\Lambda^2} \beta_{n,2}(\chi) \, ,
\ea
remembering that
\ba
\beta_k= (-1)^{k+1}\sum_{n=0}^4 \frac{n!}{(n-k)!} \alpha_n\,.
\ea

\subsection{Decoupling limit}

The decoupling limit for this theory is thus the standard one derived in (see Ref.~\cite{deRham:2010ik} for the contribution from the helicity-0 and -2 modes and Ref.~\cite{Ondo:2013wka} for the contributions from the helicity-1 modes and their mixing with the helicity-0 mode) with the addition of some extra terms. We sketch the vierbein derivation of this DL in the Appendix.
In what follows we keep the same notation as in \cite{deRham:2014zqa} where the full decoupling limit for massive gravity is given in Eq.~(8.52). We may thus write
\ba
\label{eq:DL_all}
\L_{\rm DL}=\L^{(0)}_{\rm DL}+\L^{\rm (New\ Terms)}_{\rm DL}\,,
\ea
where $\L^{(0)}_{\rm DL}$ are the usual decoupling limit terms and is given by
\ba
\label{eq:DL_0}
\L^{(0)}_{\rm DL}& =& -\frac 14 h^{\mu\nu}\hat{\mathcal{E}}^{\alpha\beta}\mn h_{\alpha \beta}
+\frac{\Lambda^3}{8}\sum_{n=0}^4 \bar{\alpha}_n  h^{\mu\nu}
\((4-n)X^{(n)}\mn+n X^{(n-1)}\mn\)  \\
&+&\L_{\rm DL}^{\rm Vectors, (0)}+\frac{h\mn}{2\mpl}T^{\mu\nu}\,,\nn
\ea
with $X^{(n)}\mn= \mathcal{E}\mathcal{E} \Pi^n$, $\Pi\mn=\p_\mu \p_\nu \pi /\Lambda^3$, where $\pi$ is the helicity-0 mode. $T^{\mu\nu}$ is the stress-energy tensor for the matter field and $\L_{\rm DL}^{\rm Vectors, (0)}$ is the usual helicity-1 contribution \cite{Ondo:2013wka}.  \\

The new contributions to the decoupling limit are given by two parts. One are purely $\pi$ interactions, and the second are helicity-0/helicity-1 interactions
\ba
\L^{\rm (New\ Terms)}_{\rm DL}=\L^{\rm (New\ Scalar\ Terms)}_{\rm DL}+\L^{\rm (New\ Scalar- Vector\ Terms)}_{\rm DL} \, .
\ea
The new scalar interactions are
\ba
\label{eq:DL_new}
\hspace{-15pt}\L^{\rm (New\ Scalar\ Terms)}_{\rm DL} &=& \frac{\mpl^2 m^4 }{4 \Lambda^2}\sum_{n=0}^4 \alpha_{n,2}(t + \dot \pi/\Lambda^3) \mathcal{E} \mathcal{E} \Pi^n \\
&=& \frac{\Lambda^4 }{4 }\sum_{n=0}^4 \alpha_{n,2}(t + \dot \pi/\Lambda^3)  \left[ (4-n) {\cal L}_n^{\rm (3d) } + n {\cal L}_{n-1}^{\rm (3d) }   \right]\\
&=&\frac{\Lambda^4 }{4 }\sum_{n=0}^4 \left[ (4-n)  \alpha_{n,2}(t + \dot \pi/\Lambda^3)  + (n+1)  \alpha_{n+1,2}(t + \dot \pi/\Lambda^3)   \right]    {\cal L}_{n}^{\rm (3d) }   \, ,
\ea
and the $\L_n^{\rm (3d)}$ are the 3d counterparts of the $\U_n$. More precisely,
\ba
\L_n^{\rm (3d)}=\mathcal{E}^{i_1\cdots i_n j_1 \cdots j_{3-n}}\mathcal{E}^{i_1'\cdots i_n'}{}_{j_1\cdots j_{3-n}} \Pi_{i_1i_1'}\cdots \Pi_{i_ni_n'}\,,
\ea
where the indices are summed over the spatial directions. The new scalar-vector interactions are given by
\ba
\label{eq:newVScoupling}
\L^{\rm (New\ Scalar- Vector\ Terms)}_{\rm DL} =  \frac{\Lambda^3}{4} \sum_{n=0}^4 \left[  n \bar \alpha_{n,1}(t+ \dot \pi/\Lambda^3) \partial^{\mu} V^{\nu} X^{(n-1)}_{\mu\nu}\right]\,,
\ea
as derived in Appendix~\ref{Appendix}.

\subsection{Families of DL theories}

Before proceeding to the stability analysis, it is worth discussing the meaning of the DL and its relation to the full exact solution. It is central to the definition of any space-time manifold that it can be split up into open charts for which in the vicinity of any point $x^{\mu}$ the metric looks as close to Minkowski as desired by reducing the size of the charts. Given a geometry whose typical curvature scale $\sqrt{R_{abcd}R^{abcd}}$ is $H^2$, and the manifold is smooth, then if we choose the charts to be of size $L \lesssim H^{-1}$, the metric inside each chart can be written as $g_{\mu\nu} = \eta_{\mu\nu}+ h_{\mu\nu}/\mpl$ where $h_{\mu\nu}/\mpl \lesssim 1$. Each of these charts centered on a given point $x^{\mu}$ can be viewed as the locally inertial frame of an observer located at $x^{\mu}$. The DL gives the leading contributions to an effective field theory which describes physics within a given chart of size $L \lesssim H^{-1}$.
\\

\begin{center}
\begin{figure}[h]
\begin{center}
 \includegraphics[trim=0cm 3cm 0cm 4cm, clip=true, width=9cm]{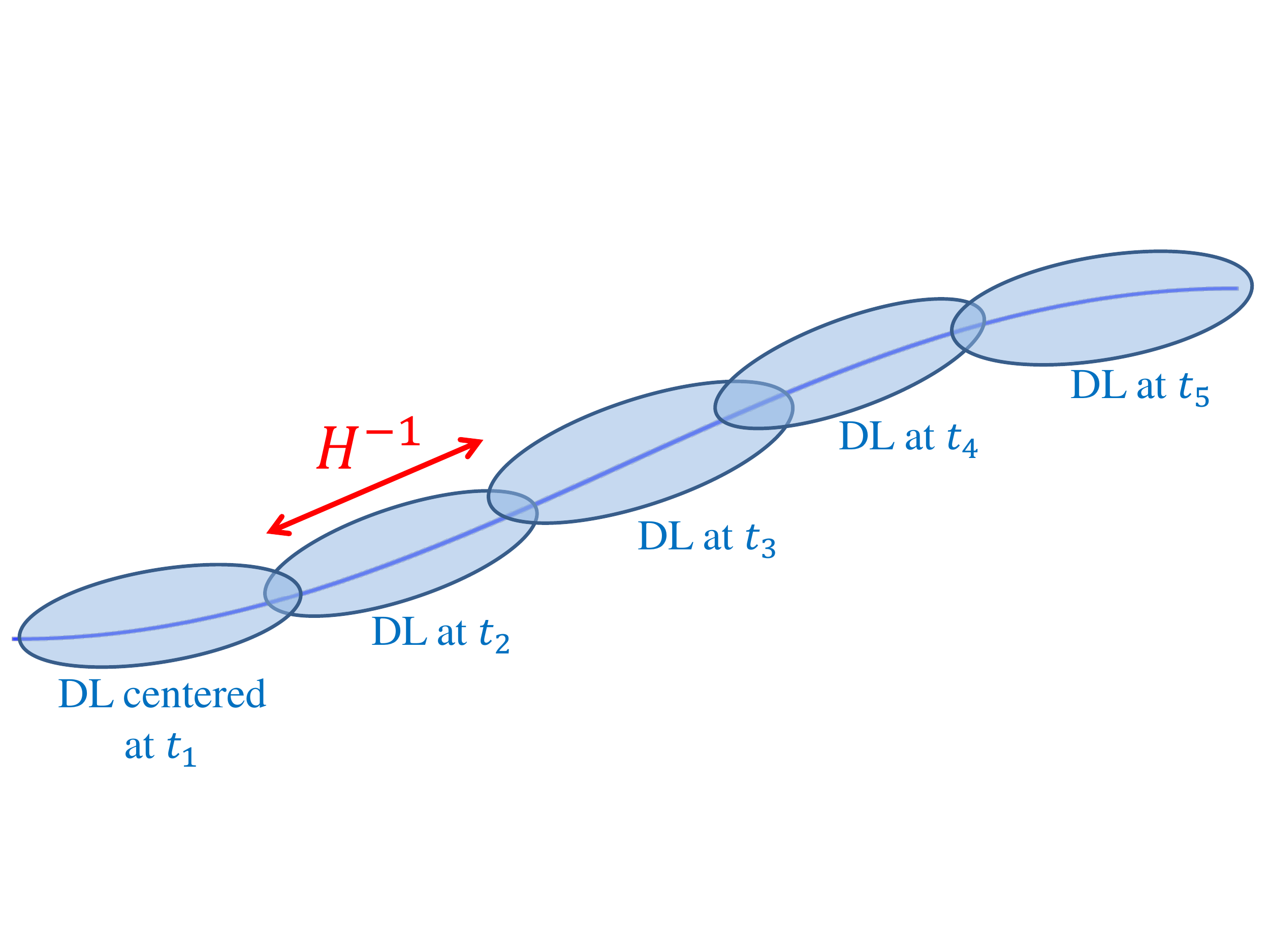}
  \caption{Families of DL theories centered around different times.}
 \label{FIG:DL_families}
 \end{center}
\end{figure}
\end{center}

In the previous section, we have derived not one but actually a family of DL theories, each centered about a specific point in space and time $x^{\mu}$, see Fig.~\ref{FIG:DL_families}.
Every single one of these DL theories is valid for one chart of size $\lesssim H^{-1}$ about the point where there are defined.  At a given time, all the DL theories look identical at any point in space, which is a simple consequence of the homogeneity assumption. The theories do however behave differently for different times. \\

This family of DL theories can therefore capture the whole history of the cosmic expansion of the Universe giving a good description of the evolution of short wavelength modes. The DL only fails to account for long wavelength modes which are bigger than their respective $H^{-1}$ because such modes are sensitive to how the different charts are patched together. The exact FLRW solutions found in these DL theories are thus valid for all time (as can be checked by comparing directly with the exact solutions of the full theory). The  stability analysis that will be presented below captures all the modes within the respective horizon of the DL theories but fails to account for the long-wavelength modes. However upon reaching these distance scales any instability which arises at the horizon scale is harmless.

\section{Stability Analysis}
\label{dec:stability}

As we emphasized in section~\ref{sec:SA}, the Friedmann equation that describes the exact FLRW solutions remains finite in the DL derived in \eqref{eq:DL_all}. This means that it should be possible to find a solution in the DL which describes the DL form of the exact FLRW solutions.

A first guess might suggest that these solutions would correspond to taking the ansatz $\pi= \pi(t)$. However this is not the case because the DL form of the metric assumes that the metric is of the form $g_{\mu \nu} = \eta_{\mu \nu} + h_{\mu \nu}/\mpl$ where $h_{\mu\nu}$ is the canonically normalized field. Thus as we have discussed above, the decoupling limit should always be taken in a gauge in which the metric is locally Minkowski. In other words the DL describes the approximate description of physics in a locally inertial frame in the vicinity of a given space-time point $x^{\mu}$ and will break down at a distance of order the curvature radius in the IR (see Fig.~\ref{FIG:DL_families}). Since any metric may be put in the locally inertial form $g_{\mu \nu} = \eta_{\mu \nu} + h_{\mu \nu}/\mpl$ by means of a diffeomorphism, there is no difficulty in describing the DL of any solution of the full theory. To make contact with the exact FLRW solutions we must then perform a coordinate transformation that takes us away from the global cosmic time slicing, to the Fermi normal coordinate system appropriate to an observer.

\subsection{From the unitary to the local Fermi normal gauge}

In this section we show how the FLRW metric in unitary gauge can be mapped onto the local Fermi normal system of coordinates and the expression for the \stu fields in that frame.
Starting with FLRW in unitary gauge appropriate to cosmic time slices (we now use capital coordinates $T,\vec X$ for cosmic time system and $t,x$ for the Fermi coordinates)
\ba
\d s^2_g  = -N^2(T)\d T^2+a^2(T)\d \vec X^2\qquad{\rm and}\qquad
\d s^2_{f}  = -\d T^2+\d \vec X^2\,.
\ea
In order to go into the Fermi normal system of coordinate, we follow the same procedure as in \cite{Nicolis:2008in,deRham:2010tw} and define the new set of coordinates $\{t, x^i\}$ with
\ba
\label{eq:Xi}
X^i&=&\frac{1}{a(T)}x^i
\equiv\phi_{\rm FN}^i(t,x)\\
T&=&\chi(t)-\frac 12 \frac{H(\chi(t))}{N(\chi(t))}x_i x^i\equiv \phi_{\rm FN}^0(t,x^i)\,,
\label{eq:T}
\ea
where the FN subscript is there as a reminder that the \stu fields $\phi^a_{\rm FN}$ are in the Fermi normal system of coordinates. The function $\chi$ is defined implicitly as $\chi'(t)N(\chi(t))=1$. \\

In the FN frame,   the dynamical FLRW metric and the Minkowski reference metrics take the form (locally, i.e. up to quadratic order in $x^i x_i=r^2$),
\ba
\d s_g^2 &=& -\(1-\(\frac{\dot{H}}{N}+H^2\)r^2\)\d t^2+\(1+H^2 r^2\)\d r^2+r ^2 \d \Omega_2^2\,\\
\d s^2_f&=&\p_\mu \phi_{\rm FN}^a\p_\nu \phi_{\rm FN}^b \eta_{ab} \d x^\mu \d x^\nu\,.
\ea
Performing the change of radial variable, $r\to R$ with $r= R\(1-H^2 R^2 + \mathcal{O}(H^4 R^4)\) $, we have
\ba
\d s_g^2 &=& -\(1-\(\frac{\dot{H}}{N}+H^2\)R^2\)\d t^2+\(1-\frac 12 H^2 R^2\)\(\d R^2+R ^2 \d \Omega_2^2\)\\
\label{eq:h_FN}
&=& \(\eta\mn+ h\mn^{\rm (FN)}\)\d x^\mu \d x^\nu
\,.
\ea
Notice however that in the DL we will scale $H\to 0$ keeping the ratio $H/m$ fixed, so that there is no difference between the coordinate $r$ and $R$ at leading order.

The \stu fields $\phi^a_{\rm FN}$ may be split into a scalar $\pi_{\rm FN}$ and vector mode $V^\mu_{\rm FN}$. We are free to fix a gauge for the vector modes and work with $V_{\rm FN}^\mu=\(V_{\rm FN}^0,0\)^T$,
\ba
\phi^a_{\rm FN}=x^a-\frac{V_{\rm FN}^a}{m \mpl}-\frac{\eta^{ab}\p_b \pi_{\rm FN}}{\Lambda^3}\,.
\ea
From the expressions \eqref{eq:Xi} and \eqref{eq:T} for the change of coordinates, we immediately infer the expression for the scalar and vector parts of the \stu fields in the Fermi normal system of coordinate,
\ba
\label{eq:pi_FN}
\pi_{\rm FN}&=&\bar\pi_0(t)+\frac{\Lambda^3}{2}\(1- \frac{1}{a(\chi(t))}\)R^2 \\
\label{eq:V_FN}
V^\mu_{\rm FN} &=& \frac 12 H(\chi(t)) m \mpl \(\frac{1}{N(\chi(t))}+\frac{1}{a(\chi(t))}\) R^2 \ \delta^\mu_0\,.
\ea
The contribution from $\bar \pi_0(t)$ satisfies
\ba
\label{eq:ddotpi0}
\frac{\ddot{\bar \pi}_0(t)}{\Lambda^3}=-1+\frac{1}{N(\chi(t))}\,.
\ea
In the decoupling limit, the Hubble parameter scales as the mass $m$, so $H m \mpl$ scales as $\Lambda^3$ and remains constant in that limit.
We see that the vectors are present already for a background FLRW metric. However they do not contribute to the modified Einstein equation in the decoupling limit since the vectors do not couple to the helicity-2 mode in that limit. Moreover we shall see later, if the parameters $\alpha_n$ or $\beta_n$ were constant,  this background profile for the vector would not affect the vectors or scalars fluctuations {\it in the decoupling limit}. However in the present case where the $\alpha_n$ or $\beta_n$ are promoted to functions of the \stu fields, one needs to take care of the contribution from the background vectors to the scalar fluctuations.

\subsection{Friedmann equation in the Decoupling Limit}

Having obtained the form of the FLRW solutions in the DL, we may now verify our assertion that the exact Friedmann equation may be recovered from the DL equations of motion.  Specifically, in the DL limit the Friedmann equation is obtained by varying the action \eqref{eq:DL_all} with respect to the $(00)$-component of the helicity-2 mode $h_{\mu\nu}$. The modified Einstein equation is given by
\ba
\mpl \delta G\mn=\frac 1\mpl T\mn+\frac{\Lambda^3}{4}\sum_{n=0}^8  \bar{\alpha}_n \left[
(4-n)X^{(n)}\mn+n X^{(n-1)}\mn\right]\,.
\ea
For the FLRW background in the Fermi normal system of coordinate, the $(00)$-component of $X^{(n)}\mn$ is given by
\ba
X^{(n)}_{00}=-\L_n^{\rm (3d)} [\Pi_{\rm FN}]= -3 ! \(1-a^{-1}\)^n\,.
\ea
The modified Friedmann equation is thus given by
\ba
3\mpl H^2 = \frac{\rho}{\mpl}-\frac 32 \Lambda^3\sum_{n=0}^4
 \bar{\alpha}_n \left[(4-n)\(1-a^{-1}\)^n+n\(1-a^{-1}\)^{n-1} \right]\,.
\ea
This is precisely the Friedmann equation derived in the full theory \eqref{eq:Friedmann}, with $\alpha_n(\chi)\to \bar{\alpha}_n$ in the DL.
So even though the DL metric is expressed locally as Minkowski plus a small correction, it is fully capable of keeping track of the physics of other backgrounds such as FLRW in this case. \\

\subsection{SVT Decomposition}

Having established that the decoupling limit successfully captures aspects of the full theory and in particular its exact FLRW solutions, we now use the DL to establish the stability of this background. We therefore consider the decoupling limit derived in section~\ref{sec:DL} with the following background and perturbation split for the helicity-2,1 and 0 modes,
\ba
&& h\mn=h\mn^{\rm (FN)}+v\mn \, ,\\
&& \pi= \pi_{\rm FN} +\delta \pi \, ,\\
&& V^\mu = V^\mu_{\rm FN}+\delta V^\mu\,,
\ea
where $\pi_{\rm FN}$ and $V^\mu_{\rm FN}$ are given in \eqref{eq:pi_FN} and \eqref{eq:V_FN} and $h\mn^{\rm (FN)}$ is given in \eqref{eq:h_FN}.\\

Despite the presence of a vector background, in the decoupling limit the latter only affects the stability analysis through the terms derived in \eqref{eq:newVScoupling}.
To be more precise, in the DL the vectors can only enter at most quadratically. Terms that involve more than two vectors enter at a scale higher than $\Lambda$ and do not survive the DL. This means that the background vector $V^0_{\rm FN}$ could only enter the DL stability analysis through these three (symbolic) types of operators,
\ba
\L^{\rm (DL)}_{V^0_{\rm FN}}\supset \bar \beta_n V^0_{\rm FN} \p \delta V \p  \delta \pi F_1(\p^2 \pi_{\rm FN})
+\bar \beta_n (V^0_{\rm FN})^2 (\p \delta \pi)^2 F_2(\p^2 \pi_{\rm FN})
+ \bar \beta_{n,1} V^0_{\rm FN}  (\p \delta \pi)^2 F_3(\p^2 \pi_{\rm FN})\,.\qquad
\ea
An explicit calculation shows that neither of the two first types of operators going as $F_{1,2}$ are present for the background considered. So the background vector only affects the DL stability analysis through terms of the form $\bar \beta_{n,1} V^0  (\p \delta \pi)^2 F_3(\p^2 \pi_{\rm FN})$, arising from \eqref{eq:newVScoupling}. Moreover these only affect the scalar fluctuations and not the vector fluctuations. This makes the stability analysis much simpler.

\subsection{Vector fluctuations}

The new mixing between the vector and the scalar arising from \eqref{eq:newVScoupling} leads to a piece linear in the vector fluctuation  $\delta V^\mu$ going symbolically as $\p \delta V\,  \p \delta \pi F(\p^2 \pi_{\rm FN})$. This new contribution can be diagonalized via the field redefinition $\delta V^\mu = \delta \tilde V^\mu + \lambda^\mu \delta \pi $  as will be performed in \eqref{eq:vecShift}. This field redefinition affects the scalar fluctuations but not the vector ones, and is thus irrelevant for the stability analysis of the vectors.

Next, the vector fluctuations enter quadratically through the terms in $\L_{\rm DL}^{\rm Vectors, (0)}$ (since we have established that the background vector $V^0_{\rm FN}$ does not contribute at second order in perturbations in the DL).
Making use of the relations (\ref{eq:pi_FN}-\ref{eq:ddotpi0}), this give
\ba
\L_{\delta V}^{(\rm DL)}= -\frac{1}{16}\left[c_1 F_{ij}F^{ij}+2 c_2 F_{0i}F^{0i}\right]\,,\quad
\ea
with $F_{\mu\nu}=\p_\mu \delta \tilde V_\nu - \p_\nu \delta \tilde V_\nu$ and
where $c_1$ and $c_2$ are functions of the scale factor and the lapse,
\ba
\label{eq:c1}
c_1 & = & \frac1a \left[\Sigma_{1,0}(a)+\frac{(a-N)}{2N}\Sigma_{2,0}(a)\right] \, , \\
\label{eq:c2}
c_2 & =&  \frac{2N}{a(a+N)} \left[  \Sigma_{1,0}(a)+\frac{(a-N)}{2N}\bar \beta_3 \right]  \, .
\ea
As a consistency check, we see that in the case where $a=N$ we recover a Lorentz invariant result, and the previous Lagrangian is proportional to $F\mn^2$ as it should be. \\

The vectors are manifestly stable for all time (i.e. for all values of the scale factor and the lapse, $a,N>0$) as long as $c_1, c_2>0$. This is  automatically satisfied if the parameters $\bar \beta_{1,2,3}$ are positive
\ba
\label{eq:CONs1}
\bar{\beta}_{1,2,3}\ge 0\,,
\ea
and as long as not all three $\bar{\beta}_k$ vanish simultaneously. By themselves these conditions are easy to satisfy from the outset.
The conditions \eqref{eq:CONs1} are stronger than necessary. 

\subsection{Scalar fluctuations}

\label{sec:ScalarFluctuations}

Next we turn to the scalar fluctuations. As usual we need to diagonalize the helicity-2 and -0 modes. In the present case we also need to diagonalize the helicity-1 and -0 modes. We perform these diagonalizations one after the other.

\subsubsection{Diagonalizing the helicity-2 and -0 modes}

This diagonalization is performed using the following field redefinition
\ba
v \mn \d x^\mu \d x^\nu  = \tilde v \mn \d x^\mu \d x^\nu 
 + \frac{1}{2a^2} \delta \pi \left[
-\(\Sigma_{1,0}+\frac{a-N}{N}\Sigma_{2,0}\)\d t^2
+\Sigma_{1,0}\d \vec{x}\,{}^2
\right]\,,
\ea
where the $\Sigma_{n,i}$ are given in  \eqref{eq:Sigma}. 
Once again we recover the standard Lorentz-invariant diagonalization expected about flat space when $a=N=1$. \\

This diagonalization induces the following coupling to matter,
\ba
\L_{\rm matter}= \frac{1}{2 \mpl}\tilde v  \mn T^{\mu\nu}
+\frac{1}{4 a^2 \mpl}\delta \pi \left[-\(\Sigma_{1,0}+\frac{a-N}{N}\Sigma_{2,0}\) \rho 
+3\Sigma_{1,0} \ p \right]\,.
\ea
The field $\delta \pi$ couples the correct way to matter at all times (i.e. for any $a$ and $N$) if
\ba
\bar{\beta}_{1,2}\ge 0 \quad {\rm and} \quad \bar{\beta}_3=0\,.
\ea
These are sufficient conditions, but of course there may be cases where these are violated and the theory is stable. 

\subsubsection{Diagonalizing the helicity-1 and -0 modes}

A new feature relative to the standard analysis is that we must also diagonalize the helicity-1/helicity-0 couplings that appear from \eqref{eq:newVScoupling}. This is most easily achieved by making use of the $U(1)$ symmetry that mixes $\partial^a \pi$ and $V^a$. We first  choose the following gauge for the vector fields
\ba
\label{eq:Gauge}
\partial_i V^i = -\frac{2}{a^2 c_2} S_{1,1}(a) \dot \pi \, ,
\ea
where  $c_2>0$ is defined in \eqref{eq:c2} and $S_{n,i}$ is defined in \eqref{eq:Sigma}.
In this gauge the de-mixing of the helicity-1 and helicity-0 modes can be achieved with the field redefinition
\ba
\label{eq:vecShift}
\delta V^0 &=&  \delta \tilde V^0 + \frac{S_{2,1}(a)}{a N c_2}  \pi \\
 \delta V^i & = & \delta \tilde V^i \, ,
\ea
which preserves the choice of gauge \eqref{eq:Gauge}.

\subsubsection{Scalar effective metric}
Having diagonalized the different modes, we can now focus on the stability of the scalar fluctuations. Their Lagrangian is given by
\ba
\L^{\rm (DL)}_{\delta \pi}= -\frac 12 Z^{\mu\nu} \p_\mu \delta \pi \p_\nu \delta \pi\,,
\ea
where the effective metric $Z^{\mu\nu}$ is diagonal with $Z^{0i}=0$ and
\ba
 Z^{00} = - \Bigg[ \frac 3{8a^2} \Sigma_{1,0}^2-\frac{3}{2a^3}\Sigma_{0,2}
+\frac 3{4a} \frac{H^2}{m^2}\Sigma_{2,0} 
  -\frac{3 H}{4  m}\frac{a+N}{a^2N^2} S_{2,1} 
-\frac{2}{a^4 c_2}\Sigma_{1,1}S_{1,1} +\frac{c_1}{c_2^2 a^4}S_{1,1}^2 \Bigg] \, ,\qquad
\ea
together with
\ba
Z^{ij}= \Bigg[\frac{3}{8a^4}\Sigma_{1,0}\(\Sigma_{1,0}+\frac 23 \frac{a-N}{N}\Sigma_{2,0}\)
-\frac{1}{2a^2N }\Sigma_{1,2} + \frac{3}{4a}\frac{H^2}{m^2}\(\Sigma_{2,0}+\frac{a-N}{3N}\Sigma_{3,0}\) \\
+\frac{\dot H}{2m^2aN }\Sigma_{2,0}
+ \frac{1}{4 a^2 N^2 c_2} S_{2,1}^2 - \frac{H }{2m a N^2 }(a+N) S_{3,1}
\Bigg]\delta^{ij}\,.\nn
\ea
The absence of ghosts requires $Z^{00}<0$ and the absence of gradient instability requires $Z^{ij}>0$. Bearing in mind the (sufficient) condition \eqref{eq:CONs1} from the vector stability the theory is free from the Higuchi ghost as long as the functions of the \stu fields satisfy the following  condition
\ba
\frac14\left[24 \beta_{0,2}''+18 \frac{\beta_{1,2}''}{a}+6 \frac{\beta_{2,2}''}{a^2}+ \frac{\beta_{3,2}''}{a^3}\right]
+\frac 34 \frac H m \frac{a+N}{a^2 N^2}S_{2,1}+\frac{2}{a^4 c_2}\Sigma_{1,1}S_{1,1}
<0\,,
\label{eq:NoGhost}
\ea
once again this is a sufficient condition, (configurations which do not satisfy that condition may still be free of the Higuchi ghost).

The absence of gradient instability further restrain the range of possibilities. A sufficient condition for the absence of such instabilities is
\ba
\frac{1}{4 }\left[6 \beta_{1,2}''+4\frac{\beta_{2,2}''}{a}+ \frac{\beta_{3,2}''}{a^2}\right]
+\frac{H (a+N)}{2m aN^2}S_{3,1}
< \frac{\dot H}{2 m^2 a N}\Sigma_{2,0}\,.\qquad
\label{eq:NoGradientInst}
\ea
Since the conditions \eqref{eq:NoGhost} and \eqref{eq:NoGradientInst} involve different combinations of the $\beta_{n,2}''$, they can be simultaneously satisfied for appropriate choices of functions. Notice as well that the Friedmann equation in the DL does not involve these functions $\beta_{n,2}$, so there is therefore a wide range of possibilities which do not affect the background evolution.

\subsection{Tensor fluctuations}

Once the field redefinition that diagonalizes the helicity-2 and helicity-0 sectors has been performed then the action for the tensor fluctuations is
\ba
\L^{(\rm DL)}_{\tilde v }=-\frac 14 \tilde v ^{\mu\nu}\hat{\mathcal{E}}^{\alpha\beta}\mn \tilde v_{\alpha \beta}+ \frac{\tilde v \mn}{2 \mpl}T^{\mu\nu}\,,
\ea
which is just the usual action for gravitational waves in GR. In particular the $\alpha_n$ or $\beta_n$ functions do not enter and so the tensors are automatically stable, with no additional constraints on the parameters $\bar \beta_n$ and the functions $\beta_{n,2}$.

\section{Discussion}
\label{sec:Discussion}

In this article we have considered the existence and stability of cosmological solutions in a simple class of `Generalized Massive Gravity' theories that have the virtue that they introduce no new degrees of freedom beyond the usual 5 of the massive graviton. Our analysis indicates that
\begin{itemize}

\item The `Generalized Massive Gravity' theories straightforwardly admit exact FLRW solutions without the need to introduce additional dark sector degrees of freedom or additional dynamical metrics.

\item The effective Friedmann equation admits self-accelerating solutions that asymptote to de Sitter, i.e. $w=-1$.

\item These solutions can easily be chosen to be completely stable, i.e. free of ghost and gradient instabilities, with a relatively mild choice on the form of the mass parameter functions which enter in the Lagrangian.

\item The Decoupling Limit consistently describes the background cosmology and the non-linear Friedmann equation can be obtained from the DL equations of motion.

\item The Decoupling Limit theory is a generalization of the usual Galileon type DL theory and allows non-Lorentz invariant Galileon interactions due to the spontaneous breaking of time translation invariance.

\end{itemize}

Although it is certainly true that the DL analysis does not capture the full stability of the entire theory, it does capture the stability correctly in the regime of validity of the DL theory. Specifically the DL ignores terms which are suppressed by higher powers of $\mpl$, so as long as these are small, they cannot affect the stability of these solutions. The fact that the Friedmann equation remains finite in the DL shows that the background evolution can be consistently described by this limit. The DL is known to break down in the IR, i.e. at large distances, however it is likely that any instability in the IR is harmless since in a local theory any such instability would arise at the scale of the Hubble radius and therefore just becomes part of the background evolution. \\

In Section \ref{sec:DL} we derived the complete form of the decoupling limit theory and demonstrated that it was equivalent to the usual DL for massive gravity with the addition of a finite number of extra terms which are dependent on the rate of change of the mass parameter functions which enter in the original Lagrangian. These additional terms are rotationally invariant but not Lorentz invariant, consistent with the breaking of time translation symmetry from the background expansion. Nevertheless they preserve the global Galileon symmetry $\pi \rightarrow \pi +v_{\mu} x^{\mu}$ which is characteristic of the DL theory. From the existence of these terms in combination with the usual Galileon interactions that arise we can infer that there is an active Vainshtein mechanism at play. Incidently the condition $\bar \beta_3=0$ and $\bar \beta_2>0$ was precisely was what derived in \cite{Berezhiani:2013dw,Berezhiani:2013dca} to ensure a stable Vainshtein mechanism. So the choice of parameters selected earlier are fully compatible with a healthy Vainshtein mechanism. As well as ensuring consistency with standard solar system and astrophysical tests of gravity, the nonlinear interactions that arise in the DL limit will play a crucial role in the growth of nonlinear structure in the Universe.

\acknowledgments

We would like to thank Gregory Gabadadze, Andrew Matas, David Pirtskhalava and Shuang--Yong Zhou for useful discussions. CdR is supported by Department of Energy grant DE-SC0009946. MF and AJT are supported by a Department of Energy Early Career Award DE-SC0010600. MF is also supported by NSF grant PHY-1068380.
The authors would like to thank the Perimeter Institute for Theoretical Physics for hospitality and support during the early stages of this work.

\appendix

\section{Vierbein derivation of the Decoupling Limit}

\label{Appendix}

We sketch the details of the vierbein derivation of the decoupling limit. Here we follow the approach and notation used in \cite{Ondo:2013wka} (see also \cite{Gabadadze:2013ria}) which introduces \stu fields for both diffeomorphisms and local Lorentz transformations. The dynamical vierbein will be denoted as $e_{\mu}^a$ and the reference vierbein as $f_{\mu}^a = \Lambda^a_{b} \d\phi^a$ where $ \Lambda^a_{b} $ are the Lorentz \stu fields. Further we will use the shorthand notation
\ba
ABCD = \epsilon_{abcd} A^a \wedge B^b \wedge C^c \wedge D^d  \, .
\ea
This notation is such that all elements commute due to the combined antisymmetry of the wedge products and the $\epsilon_{abcd} $ tensor. This means that we can usual simple algebraic manipulations such as $(A+B)^4 = A^4 + 4 A^3 B + 6 A^2 B^2+4 A B^2 + B^4$. We will thus denote the dynamical vierbein by $e$ and the reference vierbein by $f= \Lambda \d \phi$ where it is understood that the right index of $\Lambda$ contracts with the nearest Lorentz index. \\

With these notations the non-linear mass term is
\ba
S_{\rm mass} = \int \frac{m^2 \mpl^2}{ 4! 4} \sum_n \alpha_n e^{4-n} (e-f)^{n} \, ,
\ea
where the additional $4!$ arises because $\int e^4 = 4! \int \d^4 x \, {\rm det}(e)$ (in writing this we are implicitly using a Euclidean convention).
The decoupling limit is performed by writing
\ba
&& e = \eta + \frac{1}{2 \mpl} h \, ,\\
&& \d \Phi = \eta - \Pi - m \frac{d V}{\Lambda^3} \, , \\
&& \alpha_n = \bar \alpha_n + \bar \alpha_{n,1}\ m \chi + \frac{m^2}{\Lambda^2} \alpha_{n,2} \, , \\
&& \Lambda = e^{m \omega}=\eta + m  \omega + \frac{1}{2} m^2 \omega^2 + \dots\,,
\ea
where $\eta^a = \eta^a_{\mu} \d x^{\mu}$, $(\d V)^a = \d V^a$, $\Pi^a = \Pi^a_b \d x^b$, $h^a = h^a_b \d x^b$, and maintaining as usual only those terms that remain finite in the decoupling limit. Note here $\omega_{ab}= - \omega_{ba}$ as required for Lorentz transformations. This anti-symmetry implies that there will be no term arising in the DL at linear order in $\omega_{ab}$ (other than the usual $\omega \d V$ term) because it must contract with a symmetric object.

There are 4 such terms that survive in the limit $S_{\rm mass}= S_1+S_2+S_3+S_4$. These are just obtained by Taylor expanding the above forms. These are given by
\ba
S_1 = \int \frac{\Lambda^3}{ 4! 8} \sum_n  \left( (4-n) \bar \alpha_n \eta^{3-n} \Pi^{n} h + n \bar \alpha_n \eta^{4-n} \Pi^{n-1} h \right)\, ,
\ea
which is the usual DL result for scalar interactions given in \eqref{eq:DL_0},
\ba
S_2 = \int \frac{\Lambda^4}{ 4! 4} \sum_n \alpha_{n,2} \eta^{4-n} \Pi^{n} \, ,
\ea
which are the new scalar interactions found in \eqref{eq:DL_new},
\ba
S_3=\int \frac{\Lambda^3}{ 4! 4} \sum_n n \bar \alpha_{n,1} \eta^{4-n} \Pi^{n-1}  \d V \chi \, ,
\ea
which are the new vector-scalar interactions presented in \eqref{eq:newVScoupling}, and finally
\ba
S_4 &=& \frac{\Lambda^6}{ 4! 4} \sum_n \bar \alpha_n   \int \Bigg[ \eta^{4-n}\frac{(n-2)}{2} \Pi^{n-2} \left( \omega(\eta-\Pi)-\eta \frac{1}{\Lambda^3} \d V\right)^2 \nn \\
&&- \eta^{4-n} \Pi^{n-1} \left(\frac{1}{2} \omega^2(\eta-\Pi)-\frac{\omega \d V }{\Lambda^3}\right)\Bigg] \, ,
\ea
which are the usual vector-scalar interactions \cite{Ondo:2013wka} included in $\L_{\rm DL}^{\rm Vectors, (0)}$ of \eqref{eq:DL_0}. Since $\omega$ is non-dynamical, it may be integrated out, which results in the expression given in \cite{Ondo:2013wka}. On rewriting these expressions in metric language using the conversion formula
\ba
\int A B C D  = 4! \int \d^4 x \,  \mathcal{E}^{\mu\nu\rho \sigma}  \mathcal{E}_{abcd} A_{\mu}^a B_{\nu}^b C_{\rho}^c D_{\sigma}^d \, ,
\ea
we recover the formula used in section~\ref{sec:DL}.

\bibliographystyle{JHEPmodplain}
\bibliography{references}

\end{document}